\documentclass[a4paper, oneside, twocolumn, notitlepage, 10pt]{extarticle_ecoc}
\usepackage{ecoc}

\usepackage{amsmath, amssymb}
\usepackage{graphicx}
\usepackage{subfig}
\usepackage{bm}
\usepackage{cuted}
\usepackage{nonfloat}

\newcommand{\Exemph}[1]{{\it #1}}

\usepackage{xspace} % for the ``space after newcommand'' problem

\usepackage[labelfont=bf, justification=centering]{caption}

\usepackage{enumitem}
\setlist[itemize]{noitemsep, topsep=0pt, itemindent=0pt,
leftmargin=0.7cm}
\setlist[enumerate]{noitemsep, topsep=0pt, itemindent=0pt, 
leftmargin=0.7cm}

\newcommand\xqed[1]{%
\leavevmode\unskip\penalty9999 \hbox{}\nobreak\hfill
\quad\hbox{#1}}
\newcommand\demo{\xqed{$\triangle$}}

\newcommand{\define}{\triangleq}

\newcommand{\transpose}{\intercal}

\newcommand{\vect}[1]{\ensuremath{\boldsymbol{#1}}}
\newcommand{\mat}[1]{\ensuremath{\mathbf{#1}}}

\newcommand{\imag}{\ensuremath{\jmath}}

\begin{document}

\selectlanguage{english}    % Standard Language

\title{Wideband Time-Domain Digital Backpropagation via\\ Subband
Processing and Deep Learning}

\newcommand{\aff}[1]{\textsuperscript{(#1)}}

\author{
	%Christian H\"ager\textsuperscript{(1)}, 
	Christian H\"ager\aff{1,2}
	and
	Henry D.~Pfister\aff{2}
%	Alexandre Graell i Amat\aff{1}, 
%	Fredrik Br\"annstr\"om\aff{1}, 
%	Alex Alvarado\aff{2}, and
%	Erik Agrell\aff{1}
%
%	Author\textsuperscript{(1)},
%	Author\textsuperscript{(2)}, 
%	Author\textsuperscript{(3)}
}

\maketitle                  % Create title and author

\begin{strip}
  \begin{author_descr}

    \textsuperscript{(1)} Department of Electrical Engineering, Chalmers
		University, Sweden
%    \mailto{alexandre.graell@chalmers.se},
%    \mailto{fredrik.brannstrom@chalmers.se},
%    \mailto{agrell@chalmers.se} 
    %%\ \ \ \mailto{(christian.haeger, alexandre.graell, fredrik.brannstrom,
		%agrell)@chalmers.se} 

		%(Email of corresponding author mandatory)

    \textsuperscript{(2)} Department of Electrical and Computer
		Engineering, Duke University, USA, \mailto{ch303@duke.edu}

    %\mailto{alex.alvarado@ieee.org} 
		
		%(Email of other authors optional)

%    \textsuperscript{(3)} Author's full affiliation,
%    \mailto{author@authors-institution.org} (Email of other
%    authors optional)

  \end{author_descr}
\end{strip}

\setstretch{1.07}
%\renewcommand\baselinestretch{1.25}

% submit to SC3 : schemes for impairment mitigation
% digital signal handling techniques

\begin{strip}
  \begin{ecoc_abstract}
		% NOTE: Don't use a blank line here but start abstract right away
		% to avoid an extra line break (40 words, as of 2018)
		We propose a low-complexity sub-banded DSP architecture for digital
		backpropagation where the walk-off
		effect is compensated using simple delay
		elements. For a simulated $96$-Gbaud signal and $2500\,$km optical
		link, our method achieves a $2.8\,$dB SNR improvement over linear
		equalization.
%		A novel DSP architecture for wideband digital backpropagation is
%		proposed. For a $96\,$GHz signal transmitted over $25 \times
%		100\,$km, a $5\,$dB effective SNR improvement ($\approx 70$\% of
%		full-inversion DBP, $7\,$dB max improvement) over linear
%		equalization is achieved at roughly $2$--$4$ times the DSP
%		complexity.  Your 35-word abstract, which will appear in the
%		conference program, should be an explicit summary of the paper
%		that states the problem, the methods used, and the major results
%		and conclusions. It should be complementary to rather than a
%		repeat of the title.
  \end{ecoc_abstract}
\end{strip}

%\glsunset{ldpc}

\section{Introduction}

%\cite{Fougstedt2018ptl} Feasibility of time-domain DBP in mainstream
%$28$-nm ASIC processing technology with fixed-point arithmetic has
%been demonstrated. 

Real-time digital backpropagation (DBP) based on the split-step
Fourier method (SSFM) is widely considered to be impractical due to
the complexity of the chromatic dispersion (CD) steps. To address this
problem, finite impulse response (FIR) filters may be used instead of
fast Fourier transforms (FFTs) to perform time-domain CD
filtering\cite{Zhu2009, Goldfarb2009, Fougstedt2017, Fougstedt2017b,
Haeger2018ofc, Haeger2018isit, Martins2018}. Indeed, the FIR filters
can be as short as $3$ taps per SSFM step, provided that the step size
is sufficiently small (i.e., many steps are used) and the filters in
all steps are jointly optimized\cite{Haeger2018isit}. 

%These results assume relatively narrowband $10$-Gbaud signals for
%which the overall effective memory introduced by CD is low. Since the
%memory increases quadratically with the signal bandwidth, it is not
%clear if time-domain DBP can be scaled gracefully also to more
%wideband signals.  

The complexity of time-domain DBP (TD-DBP) is dominated by the total
number of CD filter taps in all steps. Recent work has focused on
relatively narrowband signals (e.g., $10$ Gbaud in\cite{Haeger2018isit} and
$20$ Gbaud in\cite{Haeger2018ofc,Fougstedt2017, Fougstedt2017b}) for
which the overall CD memory is low. Since the memory
increases quadratically with bandwidth, it is not clear if TD-DBP can
be scaled gracefully also to more wideband signals.

%Recent work on time-domain DBP has focused on characterizing the
%fixed-point requirements\cite{Fougstedt2017, Fougstedt2017b,
%Martins2018} and reduction of FIR filter lengths

%Time-domain DBP has been demonstrated in fixed-point
%arithmetic\cite{Fougstedt2017, Fougstedt2017b, Martins2018}, opening
%the way for efficient hardware implementations. 

%(5 symbol periods per $100\,$km of standard single-mode fiber (SSMF))

In this paper, we consider a $96$-Gbaud signal where the delay spread
per $100\,$km amounts to $125$ symbol periods. It is shown that TD-DBP
can still offer a good performance--complexity trade-off by leveraging
digital subband processing. In particular, the group delay difference
in different subbands can be compensated almost entirely using delay
elements. A fractional delay filter is only needed after the
last SSFM step.

%Moreover, the filter optimization is accomplished using
%deep learning\cite{Haeger2018ofc, Haeger2018isit}. 

%(i.e., all chromatic dispersion filters in the SSFM, the FD filters,
%and the analysis and synthesis filters for sub-band processing)

%One possible approach to achieve a good performance--complexity
%trade-off is through digital sub-band processing. This entails a
%potential performance loss (to to possibly uncompensated sub-band
%interference), but it also reduces the effective memory per sub-band. 

\section{Subband Processing and Related Work}

% Elliott - Handbook of DSP (describes the interpretation of DFT as
% filter bank)

%subcarrier modulation as more nonlinear tolerance compared to
%single-carrier? 

%Even if it were feasible to implement DBP with FFTs/IFFTs in a DSP,
%it would not even work due to finite precision arithmetic effects. 

%Nazarathy2012, 
%Nazarathy2014a,

% nonmaximally decimated DFT filter bank
% oversampled filter bank

% uniformly modulated FB 

% trivial prototype filter = rectangular window in time domain
% => no filter in polyphase decomposition

%Mateo2008, Mateo2009,

%or reducing the guard band duration in multi-carrier transmission
%schemes\cite{Nazarathy2014}

%, Malekiha2015

Subband processing has been previously studied for both
linear\cite{Taylor2008, Ho2009, Slim2013, Nazarathy2014}
and nonlinear\cite{Mateo2010, Ip2011, Oyama2015} impairment
compensation. The idea is to split the received signal into $N$
parallel signals using a filter bank. Assuming a bandwidth reduction
by $N$, the delay spread per subband signal is reduced by $N^2$. This
can allow for significant complexity savings.

%e.g., by reducing the FIR filter lengths when performing time-domain
%CD compensation in each subband\cite{Taylor2008, Ho2009}. 

\begin{figure}[!b]
	\centering
		\includegraphics{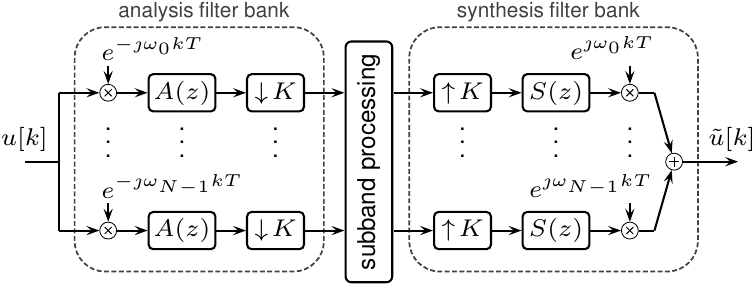}
		\includegraphics{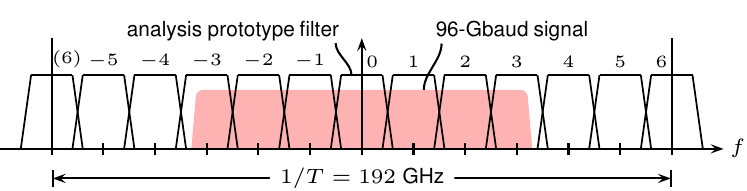}
	\caption{Filter bank (top) and signal spectrum (bottom)}
	\label{fig:filter_bank}
\end{figure}

We consider a uniformly modulated filter bank as shown in
Fig.~\ref{fig:filter_bank}. The subband signals are obtained by
filtering a downconverted version of $u[k]$ with a prototype filter
$A(z)$, where $w_i \define \frac{2 \pi i}{N T}$ is the frequency
shift. The signals are then downsampled by $K < N$ and jointly
processed.  Finally, a synthesis filter bank reassembles the signal
$\tilde{u}[k]$. Certain subbands may be inactive if they do not
contain useful signal components. Active subbands are indexed
symmetrically around the central subband according to $i \in \{-S,
\dots, S\} \define \mathcal{S}$.

\Exemph{Example 1:} Consider a $96$-Gbaud signal sampled at $1/T =
192\,$GHz. For $N = 12$ subbands, most of the spectrum falls within
the central $7$ subbands, see Fig.~\ref{fig:filter_bank}. Thus, one
may set $S = 3$. 
%Thus, setting $S = 3$ implies that $5$ out of $12$ subbands are
%inactive.
\demo

%For very
%high input powers, it may be beneficial to activate subbands $\pm 4$
%and $\pm 5$ due to spectral broadening.

%related \cite{Zhang2015a}, extension of \cite{Zhang2015} (OFC)

%Sub-band processing has also been proposed to be used in the nonlinear
%steps of the SSFM to enable larger step sizes \cite{Ip2011}. 

%\begin{figure}[!b]
%	\centering
%		\caption{Schematic representation of the signal spectrum}
%	\label{fig:signal_spectrum}
%\end{figure}

\section{Proposed DSP Architecture}

%Coupled NLSE is also known as separated-channels approach (SCA). The
%paper \cite{Leibrich2003} cites \cite{Agrawal1995} and \cite{Yu2000}
%(mean-field approach).

A theoretical foundation for DBP based on subband processing can be
obtained by inserting the split-signal assumption $u = \sum_{i} u_i$
into the NLSE. This leads to a set of coupled equations which can then
be solved numerically. Our approach is based on the SSFM proposed
in\cite{Leibrich2003}. The method is essentially equivalent to the
standard SSFM for each subband, except that all sampled intensity
signals are jointly processed with a multiple-input multiple-output
(MIMO) filter prior to the nonlinear phase rotation step. This
accounts for cross-phase modulation (XPM) between subbands but not
four-wave mixing (FWM) because no phase information is exchanged. The
method can also be used for DBP of noncoherent subband signals, e.g.,
in wavelength division multiplexing scenarios with different local
oscillators. This was done in \cite{Mateo2010}. 

%\newpage
%
%\begin{strip}
%\begin{figure*}
%	\centering
%		\includegraphics{dsp_architecture}
%	\figcaption{DSP architecture}
%	\label{fig:dsp_architecture}
%\end{figure*}
%\end{strip}

Fig.~\ref{fig:dsp_architecture} shows the proposed architecture, which
consists of the following three main components:

\begin{itemize}
	\item[(A)] Short filters compensate for %CD-induced
		pulse broadening in each subband and step 
		$\ell = 1, \dots, M$.  

		%$\ell \in \{1, \dots, M\}$.  

	\item[(B)] Delay elements are used to compensate the group delay
		difference in different subbands. 
		
%		This is different compared to the approach in \cite{Mateo2010},
%		where delay elements are only used to estimate the phase rotation
%		in the nonlinear steps, not for the actual walk-off compensation. 

%		We remark that delay elements are used only to estimate the phase
%		rotation in the nonlinear steps, not for the actual walk-off
%		compensation (which is assumed to be performed via the
%		conventional FFT/IFFT pair). (This is related to Section 3-C in
%		\cite{Mateo2008})

	\item[(C)] The MIMO filtering is performed in the time domain using
		sparse tensor decompositions.

\end{itemize}
Compared to\cite{Leibrich2003, Mateo2010}, no FFT/IFFT pairs are used.
In the following, the individual components (A)--(C) in
Fig.~\ref{fig:dsp_architecture} are described in more detail. 

%Our approach to reduce DSP complexity is \emph{not} to reduce the
%number of SSFM steps compared to\cite{Leibrich2003, Mateo2010} but to
%simplify. In fact, for time-domain processing, using many steps is
%
%Step-reducing approaches are  (and fundamentally the opposite) of
%
%We stress that our goal is not to reduce the number of SSFM steps

%It will be shown that this translates into significantly reduced
%complexity. 

%We remark that the idea of modeling the group delay difference in
%different subbands. Indeed, it is somewhat surprising that this
%approach works, because, as we will see, it puts a restriction on the
%step sizes that can be used.  

\section{(A) Pulse-Broadening FIR Filters}

%$\Dft$

%$\omega \in [-\frac{\pi}{T}, \frac{\pi}{T})$
%H(\omega + \omega_i) = 
%= e^{\imag \kappa (\omega^2 + 2 \omega \omega_i + \omega_i^2)}\\
%- 2 \kappa \omega_i = 
%by $w_i$

The frequency response of an ideal CD compensation filter is
$H(\omega) = e^{\imag \kappa \omega^2}$ where $\kappa \define
\frac{\beta_2}{2} \xi$ and $\xi$ is the propagation distance. Since
the subband signals are downconverted relative to the carrier
frequency, the filter responses have to be shifted as well. The ideal
response for subband $i$ is
\begin{align*}
	H(\omega+\omega_i) 
	%= e^{\imag \kappa (\omega + \omega_i)^2} \nonumber
= e^{\imag \kappa \omega^2}  e^{\imag \kappa 2 \omega \omega_i}
 e^{\imag \kappa \omega_i^2} 
= H(\omega)  D_i(\omega)  e^{j \phi_i},
%\label{eq:cd}
\end{align*}
where $D_i(\omega) \define e^{-\imag t_i \omega}$ with $t_i \define-
\beta_2 \xi \omega_i$ compensates the walk-off relative to the central
subband, and $\phi_i \define \beta_2 \xi \omega_i^2 / 2$. The filters
$H^{(\ell)}(z)$ correspond to $H(\omega)$ and compensate for pulse
broadening which is independent of $i$. Thus, the same filter can be
used in all subbands. Moreover, the filters are symmetric since
$H(\omega)$ is symmetric, allowing for a folded DSP implementation
with $4(L+1)$ real multiplications (RMs) assuming $2L+1$ complex
filter taps. Different filters are used in different steps (even if
the step size is the same) to avoid accumulating truncation errors
that arise from approximating $H(\omega)$ with finite-length
filters\cite{Haeger2018isit}.

%$H(\omega)$ accounts for pulse broadening, 

%The phase shift by $\phi_k$ is ignored here
%since we focus only on incoherent inter-subband processing. 

%, according to \eqref{eq:cd},

%The actual filter coefficients for $H^{(i)}(z)$ in all steps $i = 1, 2,
%\dots, K$ are jointly optimized using deep learning, as described
%below. 

\section{(B) Walk-off Delays and SSFM Step Size}

%\frac{1}{2 \pi \beta_2 \Delta f \fsub} = 
% = - \beta_2 \xi \omega_i

\newcommand{\Tsub}{T_\text{sub}}

The group delay $t_i$ depends linearly on the propagation distance
$\xi$. The step size can thus be chosen such that $t_i$ for all $i \in
\mathcal{S}$ is an integer multiple of the subband sampling interval
$\Tsub \define K T$:
\begin{align}
	%T_\text{sub} = |t_1| \implies \xi &= N K T^2 (2 \pi |\beta_2|)^{-1}  \define \delta,
	\Tsub = |t_1| \implies \xi &= \frac{N K T^2}{2 \pi
	|\beta_2| } \define \delta,
\end{align}
where $t_1$ is the group delay difference in two neighboring subbands.
Thus, as long as the step size is an integer multiple of $\delta$, the
walk-off can be compensated exactly using delay elements. 

\Exemph{Example 2:} For the parameters in Ex.~1, $K = 8$, and $\beta_2
= -21.7\,$ps$^{2}$/km, we have $\delta \approx 19.1 \,$km.  \demo

%for the considered setup (i.e., $1/T = 192\,$GHz and $N = 8$) 

%We assume $S = 8$ subbands with $2 \times$ oversampling. Thus, the
%center frequences of two adjacent subbands are spaced $\Delta f =
%8\,$ GHz apart and each subband DSP operates at $f_s = 2 \Delta f =
%16\,$ GHz. One DSP delay element $z^{-1}$ therefore corresponds to an
%implied step size of $\delta = (2 \pi \beta_2 \Delta f f_s)^{-1}
%\approx 57.4\,$km.  \demo

It is clear that the transmission distance is not necessarily an
integer multiple of $\delta$. Therefore, a fractional delay filter
$F_i(z)$ is inserted after the last step to account for any remaining
non-integer delay prior to the synthesis filter bank. 

%\begin{align}
%	[\vect{f}^{(1)}(\vect{u})]_i = u_i e^{\imag \theta_i |u_i|^2}
%\end{align}
%
%\section{Coupled NLSE}
%
%\begin{align}
%	\frac{\partial}{\partial z} \hat{u}_k = \imag \frac{\beta_2}{2}
%	\frac{\partial^2}{\partial t^2} \hat{u}_k
%	- \imag \gamma \left( 2 \sum_{q=1}^N |\hat{u}_q|^2 - |\hat{u}_m|^2 \right)
%\end{align}

\section{(C) MIMO-FIR Filter}

\begin{figure}[!t]
	\centering
		\includegraphics{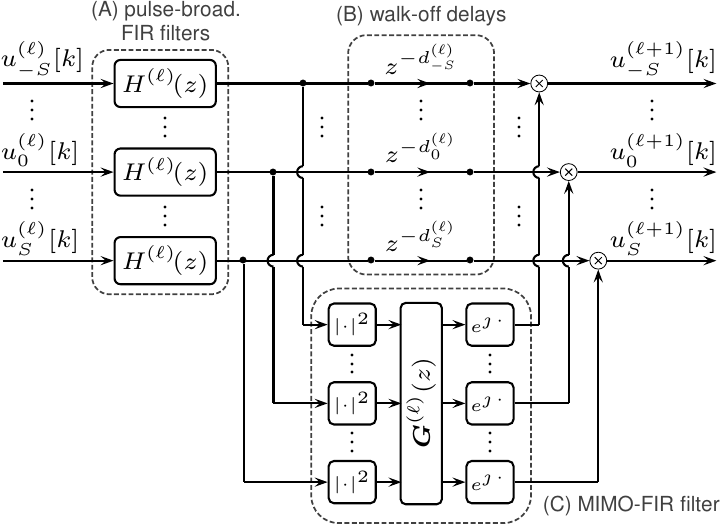}
		\caption{Proposed DSP architecture for one SSFM step}
	\label{fig:dsp_architecture}
\end{figure}

% $a_{j}^{(i)}[k]$

%(b_{-S}^{(i)}[k], \dots,
%b_{S}^{(i)}[k])^\transpose \define

%$\vect{a}^{(i)}[k] \define (a_{-S}^{(i)}[k], \dots,
%a_{S}^{(i)}[k])^\transpose$

Let $a_{i}^{(\ell)}(z)$ be the $z$-transform of the intensity signal
after the filter $H^{(\ell)}(z)$ in subband $i$ and step $\ell$ and
define $\vect{a}^{(\ell)}(z) \define (a_{-S}^{(\ell)}(z), \dots,
a_{S}^{(\ell)}(z))^\transpose$. The nonlinear phase shift is computed
based on filtered intensity signals whose $z$-transform is
\begin{align}
	 \vect{b}^{(\ell)}(z) = \mat{G}^{(\ell)}(z) \vect{a}^{(\ell)}(z), 
	%\vect{b}^{(\ell)}[k] = \mat{G}^{(\ell)}[k] * \vect{a}^{(\ell)}[k], 
\end{align}
where $\mat{G}^{(\ell)}(z)$ is an $|\mathcal{S}| \times |\mathcal{S}|$
polynomial matrix. The order $O_\ell$ of the (real and nonsymmetric)
MIMO filter, i.e., the largest polynomial degree in
$\mat{G}^{(\ell)}(z)$, is assumed to be equal to the maximum number of
walk-off delay elements in step $\ell$. 

%be equal to $d_{-S}^{(\ell)}$. 

%In order not to incur any additional processing delays due to the
%MIMO filtering, 

\Exemph{Example 3:} For the parameters as before and a step size $2
\delta$, the group delay difference of the outermost subbands
corresponds to $2 (|\mathcal{S}|-1) \Tsub = 12 \Tsub$. Thus, the order
of $\mat{G}^{(\ell)}(z)$ is $O_\ell = 12$. \demo

%\cite{Leibrich2003}
%In the previous example, the MIMO filter would have $637$ filter taps
%which would require $91$ real multiplications per subband and SSFM
%step. 
%
%$\mat{G}[k] = \mat{G}^{(s)}[k] * \dots * \mat{G}^{(1)}[k]$
%
%The number of filters increases quadratically with $S$, e.g., for $S =
%16$, one needs $16 \cdot 15 = 240$ filters. This can be reduced to one
%FFT/IFFT per subband and frequency-domain filtering. 
%
%Complexity with FFT/IFFT is 
%\begin{align}
%	\frac{n}{n-m} \left(4 \log_2(n) + 4 \right)
%\end{align}
%
%For the MIMO filter, the complexity per sub-band is
%\begin{align}
%	\left(4 \log_2(n) + 4 (2 S + 1) \right)
%\end{align}
%Assuming $n = 2^{10} = 1024$ and $S = 3$ (7 sub-bands), we get $C =
%68$ real multiplications per subband
%
%As a baseline, time domain gives $7 \times 13 = 91$ real
%multiplications per subband and SSFM step

%Transfer function (poles and zeros?)
%\begin{align}
%	\mat{H}(z) = \sum_{k=-\infty}^{\infty} \mat{H}_k z^{-k} = \prod_{i} (\mat{I} -
%	\mat{G}_i z^{-1})
%\end{align}
%
%LTI system viewpoint
%\begin{align}
%	x[k+1] &= A x[k] + B u[k] \\
%	y[k] &= C x[k] + D u[k] 
%\end{align}
%
%Compute from state-space model:
%\begin{align}
%	D, CB, CAB, CA^2B, \dots
%\end{align}
%impulse response or Markov parameters

\section{Joint Filter Optimization via Deep Learning}

``Unrolling'' all SSFM steps in Fig.~\ref{fig:dsp_architecture} leads
to a multi-layer computation graph similar to a deep neural
network\cite{Haeger2018ofc, Haeger2018isit}. Thus, joint optimization
of all filters can be achieved using tools from machine learning, in
particular deep learning via stochastic gradient descent. The tunable
parameters are 

\begin{itemize}
	\item the prototype filters $A(z)$ and $S(z)$,

	\item the filters $H^{(\ell)}(z)$ for $\ell \in \{1, \dots,
		M\}$,

	\item the MIMO filters $\vect{G}^{(\ell)}(z)$ for $\ell \in \{1,
		\dots, M\}$,

	\item the fractional delay filters $F_{i}(z)$ for $i \in
		\mathcal{S}$.
		
%		are real-valued nonsymmetric 6-tap FIR filters. They are
%		initialized to Lagrange interpolation filters. 

%		where the noninteger subband delay can be computed from the
%		total transmission distance and the SSFM step size. 

%	\item The MIMO-XPM filters are learned from scratch, i.e., they are
%		randomly initialized. To induce sparsity, a simple $L_1$-norm
%		penality (regularizer) is added to the MSE cost function. 

\end{itemize}
The optimization is performed in TensorFlow using the Adam optimizer.
The mean squared error between the transmitted and received data
symbols is used as a loss function, assuming a matched filter and
phase offset rotation after the subband processing.  Initially, the
prototype filters are raised-cosine filters, the filters
$H^{(\ell)}(z)$ are pre-optimized using least-squares
methods\cite{Haeger2018isit}, and the filters $F_{i}(z)$ are 8-tap
Lagrange interpolation filters. The MIMO filters are randomly
initialized.

A potential issue in terms of complexity is the large number of MIMO
filter coefficients, e.g., the filter in Ex.~3 is a $7 \times 7 \times
13$ tensor with $637$ real coefficients. We assume that these tensors
can be decomposed into a cascade of $F$ sparse tensors
$\mat{G}^{(F,\ell)}(z) \dots \mat{G}^{(1,\ell)}(z) =
\mat{G}^{(\ell)}(z)$, where all $\mat{G}^{(j,\ell)}(z)$ have dimension
$|\mathcal{S}| \times |\mathcal{S}|$ and order $O_\ell/F$.  To
encourage sparsity during training, we employ $L_1$-norm
regularization for all MIMO filters. 

%The (real-valued) filter coefficients are optimized further as
%described in the next section. 

%Lagrange Interpolation filter 
%\begin{align}
%	h_n = \prod_{k=0, k \neq n}^N \frac{D-k}{n-k}
%\end{align}
%$n = 0,1,\dots,N$. 

%For odd-order $N$ odd (even-length FIR filter), the delay location is
%between the two central samples. The integer delay is
%$\frac{N-1}{2}$. 

\section{Results and Discussion}

We consider a $96$-Gbaud signal (root-raised cosine, $0.1$ roll-off,
Gaussian symbols), $25 \times 100\,$km of fiber ($\alpha =
0.2\,$dB/km, $\beta_2 = -21.7\,$ps{$^2$}/km, $\gamma = 1.3\,$1/W/km),
and amplifiers with $4.5\,$dB noise figure. Forward propagation is
simulated using the standard SSFM with $6$ samples/symbols and $1000$
logarithmic steps per span (StPS). Subband TD-DBP is performed with $N
= 12$, $S =3$, $K = 8$, and a uniform step size of $2 \delta =
38.2\,$km for the first $65$ steps (see Ex.~2). The last step size is
$17\,$km for a total of $M = 66$ steps ($2.6$ StPS on average).  

\begin{figure}[!t]
	\centering
		\includegraphics{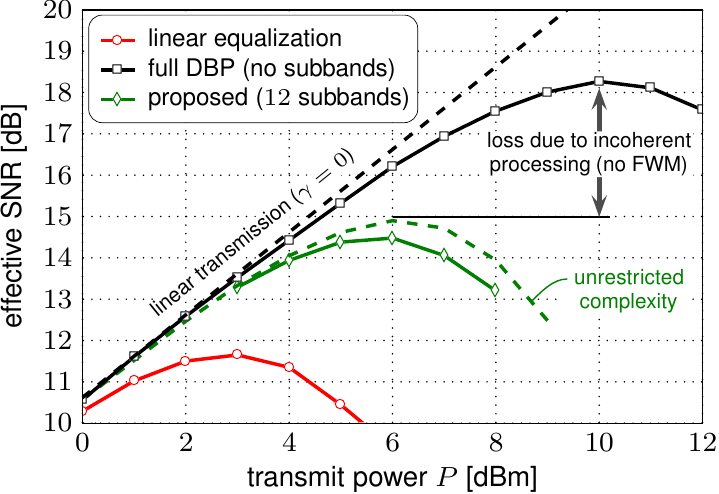}
	\caption{Simulation results}
	\label{fig:results}
\end{figure}

The results after training are shown in Fig.~\ref{fig:results}. Our
method achieves a $2.8$ dB SNR improvement over linear equalization.
The loss with respect to full DBP ($2$ samples/symbol, $1000$ StPS) is
mostly due to the incoherent subband processing. To illustrate this,
we also show results assuming essentially unrestricted complexity
(dashed line), where $S = 5$, $K = 1$, and frequency-domain filtering
according to\cite{Leibrich2003} with $1000$ StPS is used.

%Inclusion of FWM will be addressed in future work. 

%The signal squaring and phase rotation requires $6$ RMs per subband
%assuming a look-up table for the exponential function. 

%, i.e., each tensor has size $7\times 7 \times 5$

%, the number of steps is not a suitable metric
%for TD-DBP. Instead

To quantify the complexity, we use RMs focusing on the
pulse-broadening and MIMO filters which dominate the requirements. The
results were obtained with $7$-tap pulse-broadening filters ($L = 3$)
which can be implemented using $4 (L+1) = 16$ RMs. For the MIMO
filters, $F = 3$ is used. The learned coefficients were thresholded,
after which only $3812$ out of $F|\mathcal{S}|^2(O_\ell/F+1)M=48510$
total coefficients were nonzero.  This gives $3812/(|\mathcal{S}|M)
\approx 8$ RMs per subband and step on average, i.e., $24$ RMs in
total. A similar analysis for frequency-domain overlap-and-add
filtering is presented in \cite{Mateo2010}. Following the same
arguments, the number of RMs for our scenario is $4 (2 n \log_2 n + 8
n)/(n-13) \approx 98$ per subband and step with an optimized FFT size
of $n=2^7$. This is significantly more than required using TD-DBP. 

% radix-2 Cooley-Tukey: 0.5 N log2 N complex multiplications

%Both for the pulse-broadening and MIMO filters, the time-domain
%approach leads to significantly relaxed implementation requirements
%compared to utilizing FFT/IFFTs. 

%They consider essentially the same method for DBP of $12$ WDM channels
%assuming $3$ StPS. 
%
%$480$ RM per $100\,$km 
%
%we get $\approx 94$ RMs per 
%
%The number of real multiplications 
%
%Also that that the filter bank can be implemented efficiently using
%polyphase decompositions\cite{Ho2009}.  

%The (real-valued) fractional delay filters have $6$ taps and c
%
% Complexity per subband (i.e., per $24$ GHz of signal)
%
% squaring (2 real multiplications) and phase rot (4 real
% multiplications) -> 6 total
%
%9 symmetric taps, i.e., 4 complex or 16 real multiplications per
%subband and step
%
%The MIMO filter has 10 real multiplications per subband and step

\section{Conclusions}

We have proposed a novel DSP architecture for DBP based on subband
processing. Our method uses short FIR filters for the CD compensation
to achieve computational efficiency. It was shown that a proper step
size choice can significantly simplify the walk-off compensation by
using delay elements. Lastly, the complexity of the XPM MIMO filters
proposed in\cite{Leibrich2003} can be reduced by applying sparse
tensor decomposition. 

%Similar to previous work on time-domain DBP, the general approach is
%to approach relies on many SSFM steps and attempts to minimize the
%complexity per step. 
%
%Due to the fact that no phase information is exchanged between
%sub-bands, one incurs a penality compared to ideal DBP for the entire
%signal. The loss may be partially mitigated by also including FWM
%products.
%
%Future work should consider the case where phase information is
%exchanged between subbands (joint coherent processing). 

\vspace{-0.1cm}

\section{Acknowledgements}

{\footnotesize 

This work is part of a project that has received funding from the
European Union's Horizon 2020 research and innovation programme under
the Marie Sk\l{}odowska-Curie grant agreement No.~749798. The work was
also supported in part by the National Science Foundation (NSF) under
Grant No.~1609327. Any opinions, findings, recommendations, and
conclusions expressed in this material are those of the authors and do
not necessarily reflect the views of these sponsors.

}

%This work was partially funded by the Swedish Research Council under
%grant \#2011-5961. The simulations were performed in part on
%resources provided by the Swedish National Infrastructure for
%Computing (SNIC) at C3SE.} 

\vspace{-0.1cm}

\newif\iffullbib
%\fullbibtrue
\fullbibfalse

% For ECOC 2018: full paper titles mandatory
% F. M. Lastname et al., is OK

%\setcitestyle{square}

\newlength{\bibspace}
\setlength\bibspace{-0.0mm}

\newcommand{\jlt}{J.~Lightw.~Technol.}
\newcommand{\ope}{Opt.~Exp.}
\newcommand{\tit}{IEEE Trans.~Inf.~Theory}
\newcommand{\tc}{IEEE Trans.~Comm.}
\newcommand{\ofc}{Proc.~OFC}
\newcommand{\ecoc}{Proc.~ECOC}
\newcommand{\ita}{Proc.~ITA}
\newcommand{\scc}{Proc.~SCC}

\iffullbib

%{\scriptsize%
{\scriptsize
\setlength{\bibsep}{0.0ex plus 0.0ex}
\bibliographystyle{osajnl}
\bibliography{$HOME/lib/bibtex/library_mendeley}%
}%

\else

\renewcommand\baselinestretch{1.0}

\fi

%\clearpage
%
%\section{Symbols}
%
%$\ell \in \{1, \dots, M\}$ SSFM steps
%
%$\delta$ elementary step size
%
%$R_s$ baud rate (not used)
%
%$T$ original sampling interval
%
%$f_s = 1/T$ original sampling frequency
%
%$N$ number of sub-bands
%
%$K$ downsampling factor 
%
%$\rho = N/K$ oversampling factor in the sub-bands
%
%$\omega_i = i \Delta \omega$
%
%$-S, \dots, S$ active subband indices, $i$
%
%$|\mathcal{S}| = 2 S + 1$ total number of active sub-bands

\end{document}